\documentclass[letterpaper, 10 pt, conference]{ieeeconf} 

\usepackage{graphicx}
\usepackage{xcolor}
\usepackage{amsmath} %
\usepackage{amssymb}  %
\usepackage[utf8]{inputenc}
\usepackage{cite}
\usepackage{nicefrac}
\usepackage{siunitx}
\usepackage{placeins}
\usepackage{hyperref}
\usepackage[caption=false,font=footnotesize]{subfig}
\usepackage{fixltx2e}
\usepackage{nicefrac}
\usepackage{pifont}
\usepackage{siunitx}
\usepackage{verbatim}
\newcommand{\xmark}{\ding{55}}
\renewcommand{\t}[1]{\mathrm{#1}}

\newcommand{\bs}[1]{\boldsymbol{#1}}

\newcommand{\qq}{\bs{q}}
\newcommand{\dqq}{\dot{\bs{q}}}

\newcommand{\ddqq}{\ddot{\bs{q}}}

\newcommand{\vv}{\bs{v}}

\newcommand{\xx}{\bs{x}}

\newcommand{\yy}{\bs{y}}

\newcommand{\bb}{\bs{b}}
\newcommand{\cc}{\bs{c}}

\newcommand{\gggg}{\bs{g}}

\newcommand{\pp}{\bs{p}}

\newcommand{\AAA}{\bs{A}}

\newcommand{\DD}{\bs{D}}

\newcommand{\GG}{\bs{G}}

\newcommand{\KK}{\bs{K}}

\newcommand{\MM}{\bs{M}}

\newcommand{\sign}{{\mathrm{sign}}}

\newtheorem{theorem*}{Theorem}

\newtheorem{lemma*}{Lemma}

\newtheorem{corollary*}{Corollary}

\newtheorem{proof*}{Proof}

\newcommand{\q}{\boldsymbol{q}}

\newcommand{\torque}{\boldsymbol{\tau}}

\newcommand{\N}{\boldsymbol{N}}

\DeclareUnicodeCharacter{2212}{-}

\DeclareMathOperator{\atantwo}{atan2}
\newcommand*{\tran}{^\mathsf{T}}

\graphicspath{{figures}{pdftex}}

\IEEEoverridecommandlockouts                              %

\title{\LARGE \bf
Swing-Up of a Weakly Actuated Double Pendulum\\via Nonlinear Normal Modes
}

\author{Arne Sachtler$^{1,2}$, Davide Calzolari$^{1,2}$, Maximilian Raff${}^{3}$, Annika Schmidt${}^{1,2}$, Yannik P. Wotte$^{4},$\\Cosimo Della Santina$^{5,1}$, C. David Remy$^{3}$, and Alin Albu-Schäffer$^{1,2}$%
\thanks{This work was supported by the Advanced Grant M-Runners (ID:~835284) by the European Research Council (ERC).}%
\thanks{${}^{1}$German Aerospace Center (DLR), Institute of Robotics and Mechatronics; Münchener Straße 20, 82234 Oberpfaffenhofen, Germany. (\tt\small arne.sachtler@dlr.de)}%
\thanks{${}^{2}$Technical University of Munich (TUM), Department of Computer Engineering; Boltzmannstraße 3, 85748 Garching, Germany.}
\thanks{${}^{3}$University of Stuttgart, Institute for Nonlinear Mechanics; Pfaffenwaldring 9, 70569 Stuttgart, Germany.}%
\thanks{${}^{4}$University of Twente, Robotics \& Mechatronics Group; Drienerlolaan~5, 7522 NB Enschede, The Netherlands.}%
\thanks{${}^{5}$Delft University of Technology (TU Delft), Department of Cognitive Robotics; Mekelweg 2, 2628 CD Delft, The Netherlands.}%
}

\begin{document}
\bstctlcite{IEEEexample:BSTcontrol}
\maketitle
\thispagestyle{empty}
\pagestyle{empty}

\begin{abstract}
We identify the nonlinear normal modes spawning from the stable equilibrium of a double pendulum under gravity, and we establish their connection to homoclinic orbits through the unstable upright position as energy increases.
This result is exploited to devise an efficient swing-up strategy for a double pendulum with weak, saturating actuators.
Our approach involves stabilizing the system onto periodic orbits associated with the nonlinear modes while gradually injecting energy.
Since these modes are autonomous system evolutions, the required control effort for stabilization is minimal.
Even with actuator limitations of less than 1\% of the maximum gravitational torque, the proposed method accomplishes the swing-up of the double pendulum by allowing sufficient time.

\end{abstract}

\section{Introduction}
Swing-up control of a double pendulum is a very classical problem in nonlinear control usually focusing on underactuated systems such as the double pendulum on a cart~\cite{Graichen2007,Yamakita1995,Flasskamp2014}, or pendula where only the first joint (PenduBot)~\cite{Fantoni2000} or only the second joint (AcroBot) is actuated~\cite{Xin2012,Spong1995}.
However, there is also a different class of pendulum systems for which swing-up control is necessary: fully actuated systems with weak actuators.
Weak means that the motor torque limits are small compared to the system's drift - i.e., potential and inertial forces.
Beyond these torque limits, any control action will saturate.
Consider the analogy of a gymnast: the torques the gymnast can apply to the high bar are not enough to hold the body in the horizontal position.
Still, the swing-up and the stabilization to the upright position are possible.

We are especially interested in energy-efficient control approaches as these usually reduce the power drawn from the actuators, helping to succeed with weak actuation.
Several groups have dealt with energy-based approaches for swing-up of double pendula~\cite{Flasskamp2014,Fantoni2000,Xin2012,Flasskamp2017,Spong1995} and also a single pendulum~\cite{Aastroem2000}.

\begin{figure}
    \centering
    \def\svgwidth{\columnwidth}
    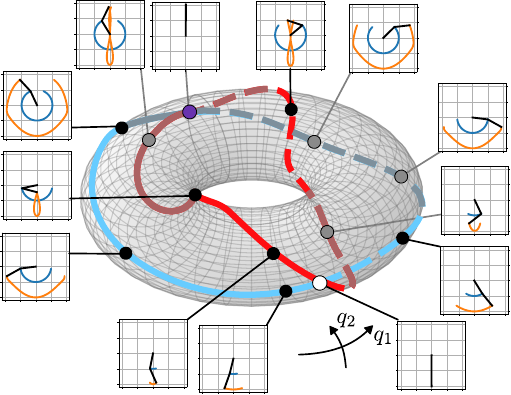
    \caption{Two pairs of nonlinear normal mode generators of the double pendulum shown on the torus. 
    Both generators (blue, red) start at the downward stable equilibrium (white dot) and meet at the upright equilibrium (purple dot) approaching homoclinic orbits. 
    A version of this illustration ironed onto the $q_1 q_2$-plane is  shown in Fig.~\ref{fig:gens}.\vspace{-.5cm}}
    \label{fig:torus_and_pendula}
\end{figure}

The double pendulum in gravity is a classical example of a chaotic system~\cite{Shinbrot1992}. 
Thus, if we just inject energy without a proper control approach, it will generally get chaotic. 
However, there exists a rich variety of periodic orbits (see~\cite{AlbuSchaeffer2023} for a classification) besides the chaos.
One class of these periodic orbits are nonlinear normal modes (NNMs), which we have studied in the past years~\cite{Kerschen2009,AlbuSchaeffer2020} and developed control approaches to stabilize, excite and mechanically implement NNMs on robots~\cite{Santina2021,Bjelonic2022,Calzolari2023,Sesselmann2021,Sachtler2022}.
NNMs can be seen as continuous families of periodic orbits of increasing energy level, each orbit having exactly two turning points.
All velocities are zero at these points, i.e., the system's energy is purely potential.
This type of orbit having two turning points is called brake orbit and the collection of all such orbits is called eigenmanifold of the mode.
The eigenmanifold encodes the system's natural dynamics in a submanifold of the state space.

The double pendulum has two NNMs that we compute using continuation methods.
Consider Fig.~\ref{fig:torus_and_pendula} for a glimpse of the results; more details and formal definitions will come later.
Both NNMs approach homoclinic orbits passing through the unstable upright position:
the two NNMs connect the downward equilibrium to the unstable one upright.
Gradually injecting energy along these NNMs can thus be used for energy-efficient swing-up to the upright position.

In simulations, we show that we can swing-up the double pendulum with weak actuators when exploiting the two NNM of the system.
Even when the motor torque limits are at $0.3\%$ of the maximal gravitational torque, we can still swing up by allowing enough time ($\approx$\qty{266}{\s}).

\section{Nonlinear Modal Analysis}
\begin{figure}[b]
    \centering
    \def\svgwidth{.60\columnwidth}
    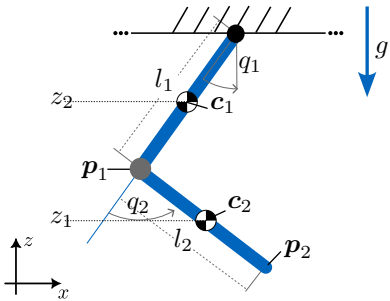
    \caption{Double pendulum under gravity. The links are assumed thin rods.}
    \label{fig:dp}
\end{figure}

Consider the thin rod model of a double pendulum in Fig.~\ref{fig:dp}. Joint coordinates are denoted by $\qq = (q_1, q_2)\in\mathbb{R}^2$.
Using the Lagrange formalism, the dynamics are
\begin{equation}\label{eq:eom}
   \MM(\qq)\ddqq + \cc(\qq,\dqq) + \gggg(\qq) = \torque,
\end{equation}
where $\MM(\qq)\in\mathbb{R}^{2 \times 2}$ is the mass matrix, $\gggg(\qq)\in\mathbb{R}^2$ are the gravitational forces, $\cc(\qq, \dqq)\in\mathbb{R}^2$ contains the Coriolis and centrifugal forces, and $\torque\in\mathbb{R}^{2}$ are the generalized forces acting at the joints (usually actuation).
The total energy is
\begin{equation}\label{eq:energy}\nonumber
    \small
    E(\qq, \dqq) = \frac{1}{2}\dqq\tran \MM(\qq) \dqq + V(\qq), \quad
    V(\qq) = V_0 + \sum_i^2 m_i g z_{i}(\qq),
\end{equation}
where $z_{i}(\qq)$ is the height of the $i$-th center of mass. 
We set $V_0$ such that $E(\qq_{\mathrm{eq}}, \boldsymbol{0}) = 0$.
As well-known, this model has a stable equilibrium at $\qq_{\mathrm{eq}} = (0, 0)$ and three unstable equilibria.
Table~\ref{tab:dp} shows the numerical values we set.

\subsection{Nonlinear Normal Modes: Brief Introduction}
Given the equations of motion~\eqref{eq:eom}, we linearize the system at the stable equilibrium $(\qq,\dqq) = (\qq_{\mathrm{eq}}, \boldsymbol{0})$
\begin{equation}\label{eq:eomlin}
    \boldsymbol{0} = \MM(\qq_{\mathrm{eq}})\ddot{\tilde{\qq}}  + \frac{\partial^2 V(\qq_{\mathrm{eq}})}{\partial \qq^2}\tilde{\qq},
\end{equation}
where $\tilde{\qq} = \qq - \qq_{\mathrm{eq}}$.
This linear system can be decomposed into two different oscillators with frequencies $\omega_1$ and $\omega_2$ evolving along the directions of the corresponding eigenvectors $\vv_1$ and $\vv_2$.
Nonlinear normal modes (NNMs) extend this concept of modes to nonlinear systems \cite{AlbuSchaeffer2020}.
However, NNMs cannot keep all of the properties of linear modes.
For example, we drop the superposition principle.

To compute NNMs, we start with the two linear modes $(\vv_1, \omega_1)$ and $(\vv_2, \omega_2)$, and investigate what happens to the nonlinear system \eqref{eq:eom}.
Let's look at $(\vv_1, \omega_1)$ first.
For very small energies the linearization approximately holds, and we see almost harmonic oscillations with frequencies $\omega_1$ along $\vv_1$ when initializing at $\qq(0) = \qq_{\mathrm{eq}} + \epsilon \vv_1$ and $\dqq(0) = \boldsymbol{0}$.
These periodic orbits will have exactly two turning points, where the velocity is zero.
The same holds for the other mode $(\vv_2, \omega_2)$.
We call this type of orbit \emph{brake orbit}.

\begin{figure}
    \centering\vspace{.3cm}
    \includegraphics[width=\columnwidth]{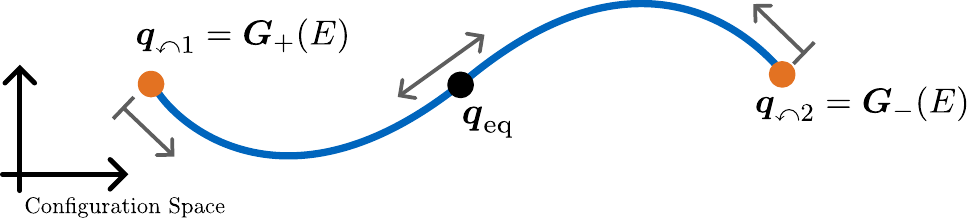}
    \caption{Sketch of one periodic orbit of an NNM for the energy level $E$ in configuration space. The system periodically oscillates between two turning points $\qq_{\curvearrowleft{}i}$ where it synchronously comes to rest before reversing direction.}\label{fig:orbitsketch}
\end{figure}

As we increase the energy, the nonlinearity will prevent these periodic orbits from extending linearly.
However, we can make them periodic motions again by adjusting the turning points slightly.
As constraints, we set that the periodic orbits must have exactly two turning points.
Adjusting the turning points is performed by multiple shooting using an integrated Jacobian of the flow~\cite{Dickinson1976}.
We repeat this process of repeatedly increasing the energy and adjusting the turning points. 
This results in a continuous family of brake orbits starting from the $i$-th linear mode.
We call this collection of orbits the \emph{$i$-th nonlinear normal mode}.

For nonlinear systems, the turning points $\qq_\curvearrowleft$ are no longer described by $\qq_\curvearrowleft = \qq_{\mathrm{eq}} \pm \epsilon \vv_i$, but by a function $\qq_\curvearrowleft = \GG_{i\pm}(E)$, which we choose to parameterize by energy $E$.
We call the images of these functions the \emph{generators} of the mode $i$.
In the nonlinear case also the period time $T_i(E)$ changes with energy.

Fig.~\ref{fig:orbitsketch} sketches one such orbit for one energy level.
The turning points are given by the generators of the mode.
By initializing the system to either $\GG_{i+}(E)$ or $\GG_{i-}(E)$ with zero velocity we obtain a periodic orbit oscillating between the two turning points.
Formally:
\begin{alignat}{4}
    \qq(0) &= \GG_{i\pm}(E) &\quad \dqq(0) &= \boldsymbol{0} \nonumber\\
    \qq(t) &= \qq(t + T_i(E)) &\quad \dqq(t) &= \dqq(t + T_i(E)).\nonumber
\end{alignat}
The turning points are at $\qq_{\curvearrowleft{}1} = \qq(kT_i(E))$ and $\qq_{\curvearrowleft{}2} = \qq(\nicefrac{1}{2}(2k+1)T_i(E))$ for $k\in\mathbb{Z}$.
Generally, the orbit does not necessarily contain $\qq_{\mathrm{eq}}$, but it is the case in our system as $\MM(\qq)$ and $V(\qq)$ are symmetric~\cite{Wotte2022,Rosenberg1966}.

Finally, when collecting all orbits of the $i$-th NNM for all energies $E$, we get a two-dimensional submanifold $\mathcal{M}_i \subset \mathbb{R}^{2n}$ of the state space. 
We call this submanifold the \emph{eigenmanifold} of the $i$-th NNM:
\begin{equation*}\label{eq:eigenmanifold}
    \begingroup\setlength\arraycolsep{1pt}
    \mathcal{M}_i = \left\{ \begin{bmatrix}\qq(t) \\ \dqq(t)\end{bmatrix} \in \mathbb{R}^{2n} \right. \left|\begin{array}{lcl}\qq(0) &=& \GG_{i\pm}(E),\\\dqq(0) &=& \boldsymbol{0},\end{array}\hfill \begin{array}{lcl}E & \in & [0, E_{\mathrm{max}}),\\t &\in& [0, T_i(E)).\end{array} \right\},
    \endgroup
\end{equation*}
where $[\qq(t), \dqq(t)]\tran$ is a solution to~\eqref{eq:eom}.
Note that it is a free choice to either use $\GG_{i+}(E)$ or $\GG_{i-}(E)$ as the generator.
They will generate the same eigenmanifold.

At this point we want to highlight that each orbit of the modes is a solution to the equations of motion.
Theoretically, following them happens naturally, without control input.

\subsection{NNMs of the Double Pendulum}
We compute the NNMs of the double pendulum, using the parameters reported in Table~\ref{tab:dp}.
The maximal potential energy $E_{\mathrm{max}} = \max_{\qq} V(\qq) = \SI{12.985}{\joule}$ is reached when the pendulum is in the upright position.
There cannot be more potential energy.
Hence, no brake orbit (and no NNM) can exist beyond this energy level $E_{\mathrm{max}}$~\cite{AlbuSchaeffer2023}.
By the continuation, we can only approach $E_{\mathrm{max}}$ but not reach it.

\begin{figure}
    \subfloat[Two pairs of NNM generators\label{fig:gens}]{
        \includegraphics[width=.48\columnwidth]{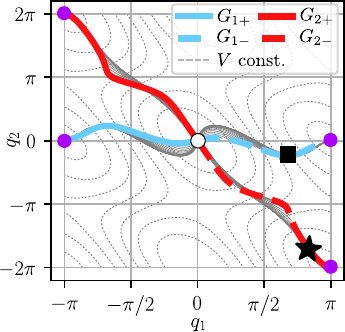}
    }
    \subfloat[Period times\label{fig:periods}]{
        \includegraphics[width=.44\columnwidth]{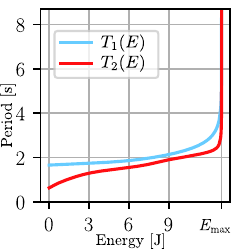}
    } 
    \\
    \subfloat[Modal oscillations for $E = \SI{11.0}{\joule}$.\label{fig:cartmodes}]{
        \includegraphics[width=\columnwidth]{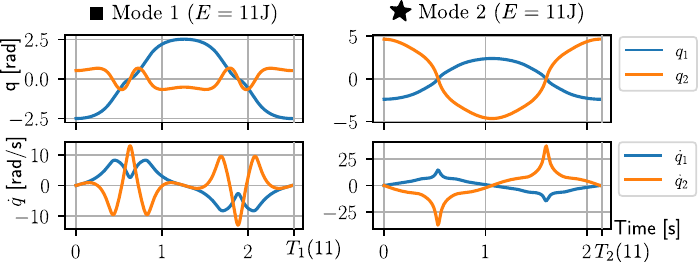}
    }
    \\
    \subfloat[Cartesian representation for some energy levels\label{fig:sos}]{
        \includegraphics[width=\columnwidth]{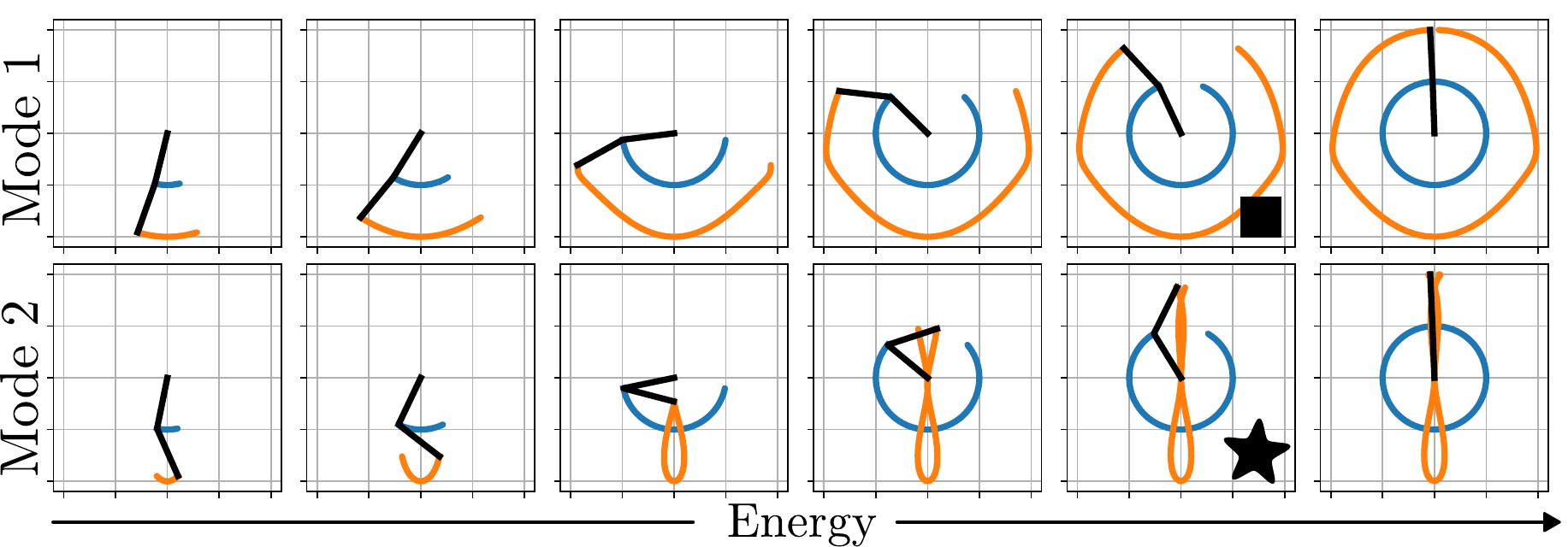}
    }
    \caption{NNMs of the double pendulum. 
    \textit{(a)} Generators. The bold and bold-dashed lines show the two pairs of generators and the thin gray lines show some trajectories of the NNMs projected into configuration space. The white/purple dots match the ones in Fig.~\ref{fig:torus_and_pendula}; 
    \textit{(b)} period times for different energy levels;
    \textit{(c)} exemplary angle evolution and angular velocity evolution for both NNMs at $E = \SI{11.0}{\joule}$. The symbols $\blacksquare$ and $\bigstar$ mark the points on the generators.
    \textit{(d)} modal oscillations for a selection of energy levels on both modes shown in Cartesian representation.}
    \vspace*{-.3cm}
    \label{fig:modes}
\end{figure}

Fig.~\ref{fig:gens} shows the two pairs of both generators $\GG_{i\pm}(E)$ (light blue, red) in configuration space, the solid gray lines show several trajectories in configuration space, and the dashed gray lines show isolines of the potential energy $V(\qq)$.
The white circle marks the stable equilibrium, and the purple dots mark the upright equilibrium.
Note, that the purple dots all correspond to the same configuration (Fig.~\ref{fig:torus_and_pendula}).
Both generators (the collection of turning points) approach the upright equilibrium as $E \rightarrow E_{\mathrm{max}}$!
We show one exemplary period of both modes in Fig.~\ref{fig:cartmodes} for $E = \SI{11.0}{\joule}$ ($\approx 85\% E_{\mathrm{max}}$).
They clearly look nonlinear.
Fig.~\ref{fig:sos} shows several orbits of both modes in Cartesian space.
The blue and orange lines show the paths of $\pp_1$ and $\pp_2$ (cmp.~\ref{fig:dp}).

Fig.~\ref{fig:periods} shows the period times $T_i(E)$.
In both cases:
\begin{equation}
    \lim\limits_{E \rightarrow E_{\mathrm{max}}} T_i(E) = \infty.\nonumber
\end{equation}

The nonlinear normal modes both approach homoclinic orbits asymptotically.
While we can never reach these homoclinics, we can get arbitrarily close.

\noindent\textbf{Main Insight:} \emph{Both NNMs connect the stable equilibrium $\qq_{\mathrm{eq}}$ to a configuration $\qq_\curvearrowleft$ arbitrarily close to the unstable equilibrium $\qq_{\mathrm{des}}$ via continuous families of periodic brake-orbits. This $\qq_\curvearrowleft$ is a turning point where all velocities vanish.}

\vspace{.2cm}

We can walk up the eigenmanifold as slowly as desired.
Finally, we will reach a periodic orbit having turning points arbitrarily close to the upright position.

\subsection{Eigenmanifold Parametrization}For the control approach we need to parameterize the eigen\-mani\-folds: we require functions $X_i$ and $\dot{X}_i: \mathcal{P} \rightarrow \mathbb{R}^n$ that map a two-dimensional parameter domain~$\mathcal{P}$ to the eigen\-mani\-fold $\forall p \in \mathcal{P}: [X_i(p), \dot{X}_i(p)]\tran \in \mathcal{M}_i$.

One of these parameters is the total energy in the system.
For the second parameter, which should encode the phase along an orbit, we look at the orbits more closely to come up with a phase.

\begin{figure}[b]
    \centering
    \def\svgwidth{\columnwidth}
    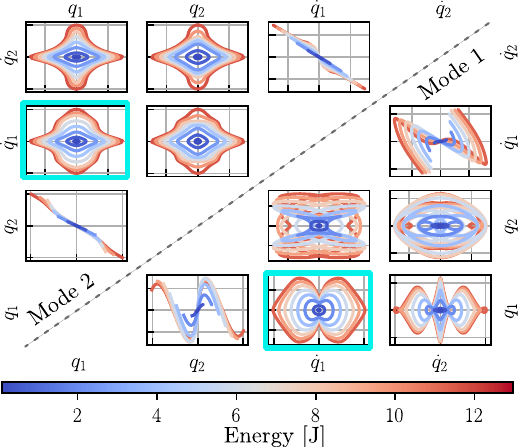
    \caption{Projection of modal trajectories onto sections of the state space. The right lower triangular matrix of plots shows mode 1 and the upper left mode 2. The color indicates the corresponding energy level. A combination of energy $E$ and the angle $\varphi = \atantwo(\dot{q}_1, q_1)$ in the $q_1\dot{q}_1$-plane (greenisch boxes) can be used to parametrize the eigenmanifolds.}
    \label{fig:allmodes}
\end{figure}

For both nonlinear normal modes, we take trajectories of different energies and project them onto two-dimensional sections of the four-dimensional state space.
There are six unique combinations of states.
We show all six projections in Fig.~\ref{fig:allmodes} for the two modes.
The color of the lines indicates the energy level.
When closely looking at the projected trajectories in Fig.~\ref{fig:allmodes}, we observe that we can use the angle \begin{equation}\label{eq:phase}
    \varphi = \atantwo(\dot{q}_1, q_1)
\end{equation}
in the $q_1 \dot{q}_1$-plane (greenish boxes) as a modal phase.
This is because there are no curls in the curves and $\varphi$ is bijective to each curve (for a fixed energy level).
Both eigenmanifolds $\mathcal{M}_i$ are parametrizable by energy $E$ and the phase $\varphi$:
\begin{equation}
   \q = X_i(E, \varphi) \quad \dqq = \dot{X}_i(E, \varphi).
\end{equation}

\begin{figure*}
    \centering
    \includegraphics[width=\textwidth]{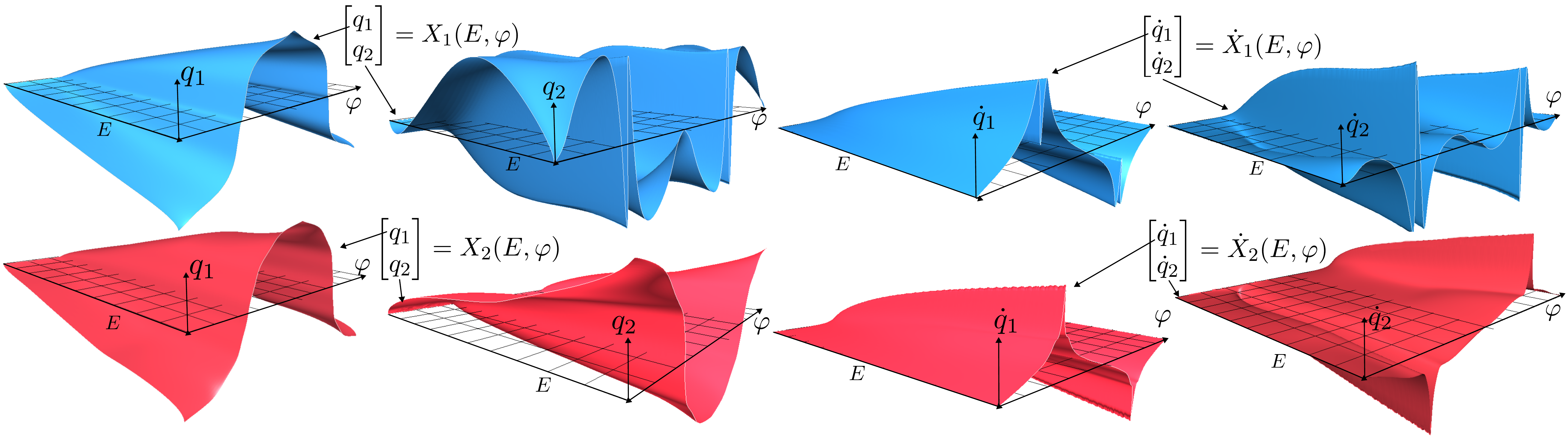}
    \caption{Eigenmanifold parametrization of both modes by the functions $X_i(E, \varphi)$ (left half) and $\dot{X}_i(E, \varphi)$ (right half). The surfaces show the respective states for given energy $E$ and modal phase $\varphi$. They are generated by the mesh-based barycentric interpolation algorithm. When taking the cross-section of the surfaces at $E=11J$ the curves in Fig.~\ref{fig:cartmodes} are obtained. Blue surfaces show mode 1 and red surface mode 2.}
    \label{fig:emm}
\end{figure*}

To obtain these functions, we compute trajectories of different energies and sample them densely in time.
As in~\cite{Bjelonic2022} this leads to a point-cloud approximation of the eigenmanifold.
We then compute energy and phase for each point and Delaunay-triangulate the point cloud in the $E\varphi$-plane using the \texttt{Qhull}-library~\cite{Barber1996}.
When given a parameter pair $(E, \varphi)$, we find the simplex containing the parameter and barycentrically interpolate~\cite{Floater2015} the states on the vertices.
Fig.~\ref{fig:emm} shows the eigenmanifold parametrization by the functions $X_i$, $\dot{X}_i$ for both modes. 
Blue surfaces show $X_1$ and $\dot{X}_1$ (mode 1) and red surfaces show $X_2$ and $\dot{X}_2$ (mode~2).

\subsection{Characteristic Multipliers}\label{sec:multipliers}

\begin{figure}[b]
    \centering
    \vspace{-.15cm}
    \includegraphics[width=\columnwidth]{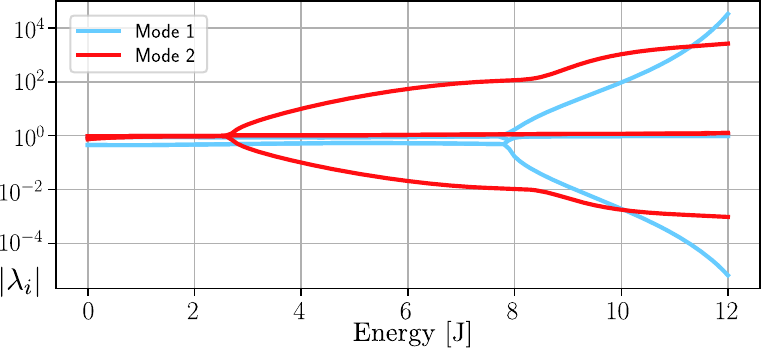}
    \vspace*{-.6cm}
    \caption{Characteristic multipliers of the orbits on the two modes. The plot shows the magnitude $|\lambda_i|$ of the eigenvalues $\lambda_i$ of the Jacobian of the Poincaré map at one of the turning points of the brake orbit.}\label{fig:floquet}
\end{figure}

We look at a \emph{conservative} double pendulum.
Hence, the NNMs will not be asymptotically stable (Liouville's Theorem~\cite{Sussman2001}).
To still get an idea of the stability of the modes, we compute the characteristic multipliers of the orbits on the two modes numerically.
Characteristic multipliers are the eigenvalues of the Jacobian of the Poincaré map~\cite{Strogatz2000} that we place at one of the turning points of the brake orbits.
This Jacobian is a four-by-four matrix computed by forward integration~\cite{Dickinson1976}.
We numerically compute the characteristic multipliers for both modes and various energy levels and show the magnitudes $|\lambda_i|$ of the four eigenvalues $\lambda_i$ in Fig.~\ref{fig:floquet}.
Note that two eigenvalues are always one due to energy conservation and the freedom of phase in a periodic motion.

The loci of the characteristic multipliers indicate different stability properties of the modes.
Mode 1 develops an unstable multiplier ($|\lambda| > 1$) at around $8J$, much later than mode 2 at $2.6J$.
For swing-up, the double pendulum will traverse the entire energy axis of Fig.~\ref{fig:floquet}.
Based on multipliers, mode 1 seems more stable in the low- and medium-energy range and is therefore more suitable for swing-up control; this remains to be validated in the experiments.

\section{Swing-Up via Nonlinear Normal Modes}Our control goal is to bring the double pendulum from the stable equilibrium to the upright position $\qq_{\mathrm{des}} = (\pi, 0)$, and stabilize it there.
Overall, we have a set of different control laws and a finite state machine decides which one to use.
The main point of this paper is the swing-up control law~$\torque_{\mathrm{su}}$, which brings the double pendulum arbitrarily close to the upright position:
\begin{equation}\label{eq:swingup}
    \torque_{\mathrm{su}} = \torque_{\mathcal{M}} + \torque_E.
\end{equation}
As in~\cite{Bjelonic2022,Santina2021}, this controller is split up into an eigenmanifold stabilizer $\torque_{\mathcal{M}}$ and an energy controller $\torque_E$.
The two components are explained in the sections \ref{subsec:emm} and \ref{subsec:energy}.

Besides the swing-up controller, we also need a controller to stabilize the upright position locally and one additional control law to inject some initial energy. 
These controllers and the state machine are summarized in \ref{subsec:statemachine}.

\subsection{Assumptions}
We assume that we have access to the full state $\qq$ and $\dqq$, and that we can evaluate the functions $\MM(\qq)$, $E(\qq, \dqq)$, $X(E, \varphi)$ and $\dot{X}(E, \varphi)$. 
We assume full actuation and torque saturation, i.e, $|\tau_i| \le \tau_{\mathrm{max}}$, where motors are weak compared to the internal model forces.
The maximum motor torque is $\tau_{\mathrm{max}}$ for both joints, but this can easily be extended for different maximal torques in each motor.

\subsection{Eigenmanifold Stabilization}\label{subsec:emm}
The goal of the eigenmanifold stabilizer is to stabilize the double pendulum onto one of the eigenmanifolds $\mathcal{M}_i$.
As in~\cite{Santina2021}, we use the parametrization of the eigenmanifold.
The functions $X_i(E, \varphi)$ and $\dot{X}_i(E, \varphi)$ are evaluated for the measured state.
Note, that $E(\qq, \dqq)$ as well as the phase $\varphi$~\eqref{eq:phase} are both functions of the state $(\qq, \dqq)$.
Therefore, we can chain the computations into single functions $Y$ and $\dot{Y}$: \begin{align}
    \qq_d &= X_i(E(\qq, \dqq), \atantwo(\dot{q}_1, q_1)) = Y_i(\qq, \dqq),\\ 
    \dqq_d &= \dot{X}_i(E(\qq, \dqq), \atantwo(\dot{q}_1, q_1)) = \dot{Y}_i(\qq, \dqq).
\end{align}
where by construction $(\qq_d, \dqq_d) \in \mathcal{M}_i$.
The functions $Y$, $\dot{Y}$ take a measured state and provide a desired state on the eigenmanifold.
We then design the eigenmanifold stabilizer \begin{equation}\label{eq:eigenmanifoldstabilizer}\nonumber
    \torque_{\mathcal{M}} = \MM(\qq)\left[-k_p \left(\qq - Y_i(\qq, \dqq)\right) - k_d \left(\dqq - \dot{Y}_i(\qq, \dqq)\right)\right],
\end{equation}
where $k_p$ is a scalar gain and $k_d= 2\sqrt{k_p}$.

\subsection{Energy Injection}\label{subsec:energy}
We want to swing up the pendulum to periodic orbits of higher and higher energy until we reach a configuration close to the upright position.
The eigenmanifold stabilizer~$\torque_{\mathcal{M}}$ has no preference on the energy level and leaves the energy coordinate uncontrolled.
We need an additional control law $\torque_E$ to regulate energy.

It is crucial to find $\torque_E$ such that the sum $\torque_{\mathcal{M}} + \torque_E$ stays within the motor torque limits:
we first compute a sliding-mode controller $\bar{\torque}$ that does not consider the torque limits
\begin{equation}
    \bar{\torque} = \sign\left(E_{\mathrm{des}} - E(\qq, \dqq)\right) \MM(\qq) \dqq \;.
\end{equation}
Since $\torque_{\mathcal{M}}(\qq, \dqq)$ is known, we can find a scaling factor $\alpha$ such that $\torque_{\mathcal{M}} + \alpha \bar{\torque}$ is within the torque limits.
When $\torque_{\mathcal{M}}$ already saturates the motors, we immediately set $\alpha=0$, i.e., no energy injection happens at this time step.
To enable fast swing-up, $\alpha$ should be as large as possible. 
Compactly
\begin{align}\nonumber
   \text{maximize} \quad & \alpha \\\nonumber
   \text{subject to} \quad & |\torque_{\mathcal{M}} + \alpha \bar{\torque}| \leq \torque_{\mathrm{max}},
\end{align}
where $\torque_{\mathrm{max}}$ is the vector of the maximum motor torques per motor.
We can rewrite the constraints to
\begin{equation}\nonumber\begin{bmatrix}
    \bar{\torque}\\
    -\bar{\torque}
   \end{bmatrix} \alpha = \xx \alpha \leq \yy = \begin{bmatrix}
    \torque_{\mathrm{max}} - \torque_{\mathcal{M}}\\
    \torque_{\mathrm{max}} + \torque_{\mathcal{M}}
   \end{bmatrix},
\end{equation}
which now reads as linear program (LP) of the form: maximize $\cc\tran \xx$ subject to $\AAA\xx  \leq \bb$.
It can be solved using a standard LP-solver.
However, as $\alpha$ is only a scalar we can find a solution analytically: 
we replace the inequality by an equality and solve each row in $x_i \alpha = y_i$ for $\alpha_i$ and then take the smallest non-negative $\alpha_i$ as the solution $\alpha_{\mathrm{opt}}$.

Finally, we set the energy controller to
\begin{equation}\label{eq:energyctrl}
    \torque_{E} = \alpha_{\mathrm{opt}} \tanh(E_{\mathrm{des}} - E) \MM(\qq) \dqq,
\end{equation}
where the $\tanh$ function is used to smoothen the $\sign$ function.
Note that $\alpha_{\mathrm{opt}}$ must be evaluated in every cycle.

\subsection{Controller State Machine}\label{subsec:statemachine}

The overall control approach bringing the double pendulum to the upright position is realized as a finite state machine traversing through the states 
\begin{equation}\nonumber
 \underset{\torque_{\mathrm{bs}}\;\text{\eqref{eq:bootstrap}}}{(\textsc{Bootstrap})} \rightarrow \underset{\torque_{\textsc{Start}}\;\text{\eqref{eq:start}}}{\vphantom{()}\textsc{Start}} \rightarrow \underset{\torque_{\mathrm{su}}\;\text{\eqref{eq:swingup}}}{\vphantom{()}\textsc{SwingUp}} \rightarrow \underset{\torque_{\textsc{Hold}}\;\text{\eqref{eq:regulation}}}{\vphantom{()}\textsc{Hold}}.
\end{equation}

\subsubsection{State \textsc{Bootstrap}}
This state is only needed to globalize the controller.
It helps to recover from any initial state, failed swing-up attempts, and/or external impacts.
The goal is to dissipate all the energy by applying an artificial friction \begin{equation}
    \torque_{\mathrm{bs}} = -\DD_{\mathrm{bs}}\dqq\label{eq:bootstrap}
\end{equation} $(\DD_{\mathrm{bs}} \succ 0)$ until reaching the stable (downward) equilibrium.

\subsubsection{State \textsc{Start}}
Suppose we are at the stable equilibrium $\qq_{\mathrm{eq}}$.
The energy controller~\eqref{eq:energyctrl} can only inject energy when $\dqq \ne \boldsymbol{0}$.
We need some initial energy to start the swing-up and accelerate along the eigenvector $\vv_i$ of the $i$-th mode of the linearized system \eqref{eq:eomlin}:
\begin{equation}\label{eq:start}
    \torque_{\textsc{Start}} = \beta\MM(\qq) \vv_i,
\end{equation}
where we set $\beta = 0.01$.
This state is active for one timestep only, and we switch to \textsc{SwingUp} immediately after.

\subsubsection{State \textsc{SwingUp}}
This is the main point of this paper.
The control law $\torque_{\mathrm{su}}$ \eqref{eq:swingup} for this state was developed in the past sections.
The goal is to bring the double pendulum to a periodic orbit having turning points~$\qq_\curvearrowleft$ arbitrarily close to the upright unstable equilibrium $\qq_{\mathrm{des}}$.
We set $E_{\mathrm{des}} = E_{\mathrm{max}}-\epsilon$ in~\eqref{eq:energyctrl}, where~$\epsilon$ is computed based on the actuator torque limits.

\subsubsection{State \textsc{Hold}}
When reaching the energy level $E_{\mathrm{des}}$ and we are at a turning point $\qq_\curvearrowleft$ we switch to a simple PD controller to stabilize the upright position $\qq_{\mathrm{des}}$:
\begin{equation}\label{eq:regulation}
    \torque_{\textsc{Hold}} = -\KK_R \left[\qq - \qq_{\mathrm{des}}\right] - \DD_{R}\dqq + \gggg(\qq)
\end{equation}
where $\KK_R$ is again a positive-definite gain matrix and \begin{equation}
    \DD_R = \zeta_R \left(\KK^{\sqrt{\cdot}}_R \MM^{\sqrt{\cdot}} + \MM^{\sqrt{\cdot}}\KK^{\sqrt{\cdot}}_R\right)
\end{equation}
is a damping matrix designed to match a damping ratio of $\zeta_R=1.0$.
The matrices $\KK^{\sqrt{\cdot}}_R$ and $\MM^{\sqrt{\cdot}}(\qq)$ are the matrix square roots of $\KK_R$ and $\MM(\qq)$, respectively.

\section{Results}\begin{figure*}
    \hspace{-.3cm}\subfloat[Experiment 1 (Mode 1): $\tau_{\mathrm{max}} = \qty{0.5}{\N\m}$ ($7.7\%\;\hat{\tau}_g$)\label{fig:q_dq_tau_10percent}]{
        \def\svgwidth{1.02\columnwidth}
        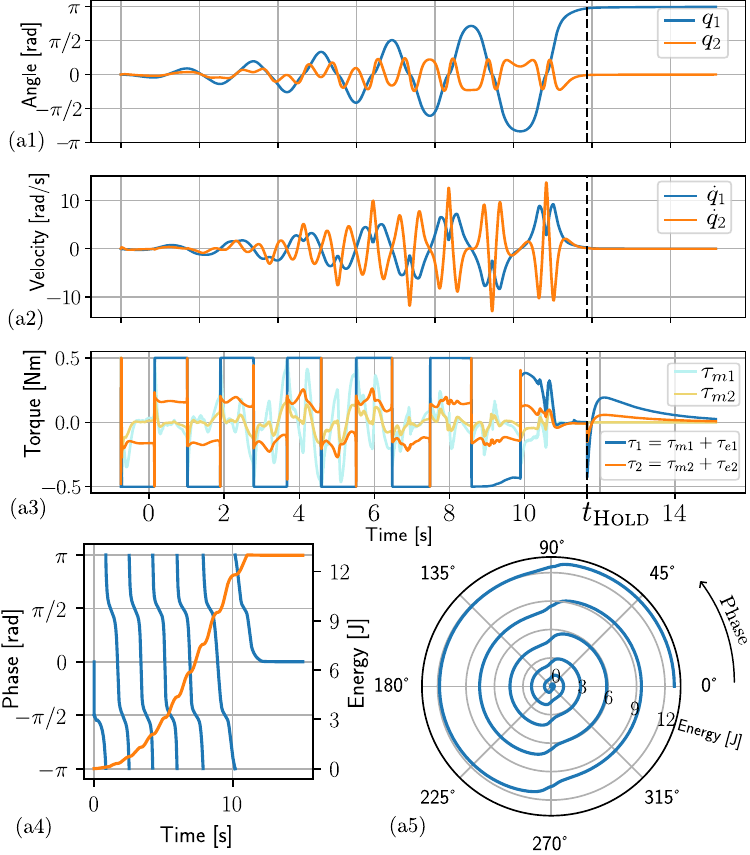
    }
    \subfloat[Experiment 2 (Mode 2): $\tau_{\mathrm{max}} = \qty{0.2}{\N\m}$ ($3.0\%\;\hat{\tau}_g$)\label{fig:q_dq_tau_02mn}]{
        \def\svgwidth{1.02\columnwidth}
        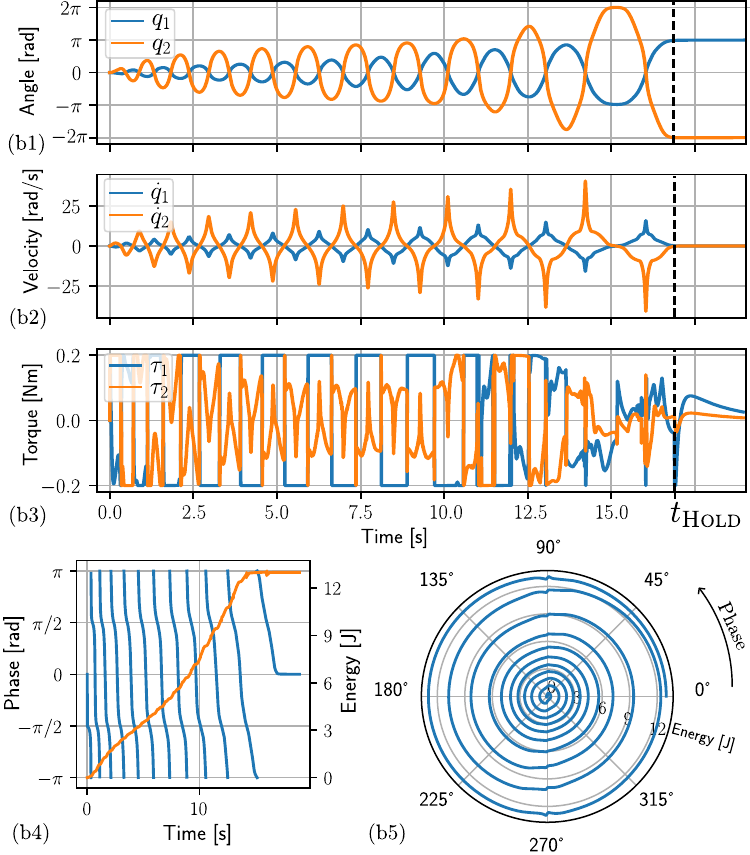
    }
    \caption{Two swing-up experiments. 
    \textit{(a)}: Swing-up via mode 1 for a motor torque limit of \qty{0.5}{\N\m} ($7.7\%$ of $\hat{\tau}_g$) and critical angle at $q_{\mathrm{crit},1} = \ang{175.58}$; 
    \textit{(b)}: Swing-up via mode 2 for a motor torque limit of \qty{0.2}{\N\m} ($3.0\%$ of $\hat{\tau}_g$) and critical angle at $q_{\mathrm{crit},1} = \ang{178.23}$.
    The top three panes \textit{(x1)-(x3)} show the joint angles, joint velocities and controller torques over time. 
    In \textit{(a3)} we additionally show the contribution of the eigenmanifold stabilization controller $\torque_{m}$. 
    Panes \textit{(x4)} shows the modal phase $\varphi$ and energy over time and \textit{(x5)} displays modal phase and energy in polar representation. 
    }\label{fig:experiments}
\end{figure*}
We evaluate the swing-up control of the weakly actuated double pendulum on a simulated double pendulum in \mbox{\textsc{MuJoCo}~\cite{Todorov2012}}.
The simulator uses a time step of $\Delta t = \SI{1}{\milli\second}$.
We experiment with the two different modes and different settings of the motor torque limit $\tau_{\mathrm{max}}$.

\begin{table}[b]
    \centering
    \caption{Parameters of the Double Pendulum}\label{tab:dp}
    \begin{tabular}{cc|c}
        \textbf{Parameter} & & \textbf{Value} \\\hline
        Link Lengths & $l_1$, $l_2$ & $\SI{0.5}{\meter}$ \\
        Link Masses & $m_1$, $m_2$ & $\SI{661.83}{\gram}$ \\
        Gravity Constant & $g$ & $\SI{9.81}{\meter\per\second\squared}$ \\
        Inertia at COM & $I$ & $\SI{0.0153}{\kilogram\meter\squared}$ \\
    \end{tabular}
\end{table}

For the model parameters presented in Table~\ref{tab:dp}, the highest gravitational torque occurs when the pendulum is stretched out horizontally $\qq_{\mathrm{horiz}} = (\nicefrac{\pi}{2}, 0)$.
\begin{equation}
    \gggg(\pm\qq_{\mathrm{horiz}}) = \begin{bmatrix}
        \mp{}6.49 \\ \mp{}1.62
    \end{bmatrix} \si{\newton\meter}
\end{equation}
We refer to the maximum as $\hat{\tau}_g = \SI{6.49}{\newton\meter}$.
This will be our reference to compare actuator torques.

We always start the swing-up from the stable equilibrium $\qq_{\mathrm{eq}}$ and consider the stabilization to the upright position successful when the pendulum reaches $||\dqq|| < \qty{e-3}{\radian\per\second}$ and $||\qq - \qq_{\mathrm{des}}|| < \qty{e-3}{\radian}$ and call the respective time $t_{\mathrm{end}}$.

The controller starts in the \textsc{Start} state and switches to the \textsc{SwingUp} state after the first time step.
Given the motor torque limit $\tau_{\mathrm{max}}$, we compute the energy level $E_{\mathrm{des}}$ that the swing-up controller must reach before the regulation controller \eqref{eq:regulation} controller can take over.
This is computed by finding the smallest energy level $E_{\mathrm{crit}}$ such that $||\gggg\left(G_{i\pm}(E_{\mathrm{crit}})\right)||_{\infty} \le \tau_{\mathrm{max}}$ and $G_{i\pm}(E_{\mathrm{crit}})_1 \ge \ang{90}$, i.e., at the turning point of the brake orbit for the desired energy level the motors must be strong enough to counteract gravity to move the pendulum to the upright position.
We call the corresponding turning point $\qq_{\mathrm{crit}} = G_{i\pm}(E_{\mathrm{crit}})$.

The first experiment is on the swing-up via mode 1.
We set the motor torque limit to \qty{0.5}{\text{Nm}}, which corresponds to $7.7\%$ of the maximal gravitational torque $\hat{\tau}_g$.
The critical angle for $q_1$ is at \ang{175.58}, i.e. the turning point where the regulation controller to the upright position can take over is at $q_1=\ang{175.58}$.
The controller switches to \textsc{Hold} at $t_{\mathrm{\textsc{Hold}}} = \qty{11.8}{\s}$ and the total swing-up time is $t_{\mathrm{end}} = \qty{15.1}{\s}$.

Panes \textit{(a1)} and \textit({a2}) in Fig.~\ref{fig:q_dq_tau_10percent} show how the joint angles and velocities evolve over time.
Note the strong nonlinearity in the oscillations.
On the bottom in \textit({a4}) and \textit({a5}) in Fig.~\ref{fig:q_dq_tau_10percent} we additionally show the energy $E$ and modal phase $\varphi$.
In contrast to harmonic oscillations, the phase does not evolve at a constant rate (frequency) over time.
Pane \textit{(a3)} shows the total applied motor torques in blue and orange; and the contribution to the total controller torque by eigenmanifold stabilization controller $\torque_{\mathcal{M}}$ in light colors.
Hence, the distance between the two lines is the torque due to the energy injection controller $\torque_{E}$.
We can observe that until we reach the critical energy level $E_{\mathrm{crit}}$, one of the motor torques always saturates if $\dqq \neq \boldsymbol{0}$.
This is because the energy injection controller aims at using all the remaining torque margin to inject energy still ensuring the control action by the eigenmanifold stabilizer is not altered.

For the next experiment, we reduce the motor torque limit to \qty{0.2}{\N\m} ($3.0\%$ of $\hat{\tau}_g$) and $q_{\mathrm{crit}1} = \ang{178.23}$ and chose mode 2 for the swing-up.
Fig.~\ref{fig:q_dq_tau_02mn} shows the results.
This time we obtain $t_{\mathrm{\textsc{Hold}}} = \qty{16.9}{\s}$ and $t_{\mathrm{end}} = \qty{18.9}{\s}$.

We repeat the experiments for various settings of $\tau_{\mathrm{max}}$ and show the results in Table~\ref{tab:times} for both modes.
For each motor torque limit $\tau_{\mathrm{max}}$ we compute the critical angle $q_{\mathrm{crit}1}$ and the percentage of $\tau_{\mathrm{max}}$ of the maximal gravitational torque~$\hat{\tau}_g$.
These values help to intuitively understand the magnitude of the motor torques.
For example, for $\tau_{\mathrm{max}} = \qty{0.02}{\N\m}$ the motor torques are not strong enough to hold the pendulum when the first angle $q_1 < \ang{179.82}$.
We believe these values should emphasize how weak the motors are compared to the internal model forces.

As expected, the swing-up times $t_{\mathrm{end}}$ get longer as the motor torques get weaker.
Swing-up via mode 2 is generally faster than via mode 1.
The period times on mode 2 are shorter and joint velocities $\dqq$ are overall higher.
Hence, the energy injection can inject more power via the faster mode.

An \xmark{} in the swing-up time column indicates that the respective setting failed.
We consider a swing-up failed when 
the double pendulum's state is too far from the eigenmanifols; 
or when energy suddenly drops; 
or when energy stagnates before completion of the swing-up.
This is the case for very weak actuators and when using mode 2 for swing-up.
The characteristic multipliers (Sec.~\ref{sec:multipliers}) indicate that mode 2 develops an unstable eigenvalue already at quite a low energy, which makes it harder to stabilize.
Due to the very weak actuation, the available forces are not sufficient to stabilize the mode.

Both modes have exhibit unstable multipliers at energy levels close to $E_{\mathrm{crit}}$.
This explains why the eigenmanifold stabilizer has significant contributions to the total torque at high energies (Fig.~\ref{fig:q_dq_tau_10percent} for $t>10s$ and Fig.~\ref{fig:q_dq_tau_02mn} for $t>12.5s$).

\begin{table}
    \centering
    \caption{Swing up times}\label{tab:times}
    \begin{tabular}{c|c|c|c|c}
        Max. Torque & Perentage & Critical Angle & Mode 1 & Mode 2\\
        $\tau_{\mathrm{max}}$ & $\nicefrac{\tau_{\mathrm{max}}}{\hat{\tau}_g}$ & $q_{\mathrm{crit}1}$ & $t_{\mathrm{end}}$ & $t_{\mathrm{end}}$ \\\hline
        \hline
        \qty{0.5}{\newton\meter} & 7.7\% & \ang{175.58} & \qty{15.07}{\s} & \qty{8.67}{\s}\\
        \qty{0.3}{\N\m} & 4.6\% & \ang{177.35} & \qty{23.03}{\s} & \qty{13.03}{\s}\\
        \qty{0.2}{\N\m} & 3.0\% & \ang{178.23} & \qty{30.47}{\s} & \qty{18.94}{\s}\\
        \qty{0.1}{\N\m} & 1.5\% & \ang{179.11} & \qty{60.774}{\s} & \xmark \\
        \qty{0.05}{\N\m} & 0.8\% & \ang{179.55} & \qty{124.05}{\s} & \xmark \\
        \qty{0.02}{\N\m} & 0.3\% & \ang{179.82} & \qty{265.91}{\s} & \xmark \\
    \end{tabular}
\end{table}

\section{Conclusions}We have shown that we can swing up the double pendulum in gravity with weak actuators thanks to the exploitation of its nonlinear modes.
Weak means that the motor torque limits are small compared to the internal model forces.
Our contributions are the nonlinear modal analysis of the double pendulum model for high energy levels, the insight that they both approach a homoclinic orbit and the design of an energy-injection controller using the available torque margin to swing up the double pendulum.

Additionally, we have shown the approach for a \emph{specific} double pendulum.
The question if we always obtain modes approaching homoclinic orbits through the upright position for arbitrary mass distribution and lengths remains open.
The symmetry of the potential and mass tensor is independent of the mass- and length distribution, so also the NNM generators will be symmetric.
This fact and the topological constraints of the torus ensure that the generators will approach homoclinics, if they exist at that energy level.
The latter is the open question.

For future work, it remains to validate and replicate the approach on a real system. 
This entails that we do not have access to the full state and the mathematical model assumptions will deviate from the real hardware. 
This is especially true for friction and other dissipative effects, which will limit the lowest admissible torque saturation level.

\bibliographystyle{IEEEtran}   
\bibliography{bibliography,ieee}

\begin{thebibliography}{10}
\providecommand{\url}[1]{#1}
\csname url@samestyle\endcsname
\providecommand{\newblock}{\relax}
\providecommand{\bibinfo}[2]{#2}
\providecommand{\BIBentrySTDinterwordspacing}{\spaceskip=0pt\relax}
\providecommand{\BIBentryALTinterwordstretchfactor}{4}
\providecommand{\BIBentryALTinterwordspacing}{\spaceskip=\fontdimen2\font plus
\BIBentryALTinterwordstretchfactor\fontdimen3\font minus \fontdimen4\font\relax}
\providecommand{\BIBforeignlanguage}[2]{{%
\expandafter\ifx\csname l@#1\endcsname\relax
\typeout{** WARNING: IEEEtran.bst: No hyphenation pattern has been}%
\typeout{** loaded for the language `#1'. Using the pattern for}%
\typeout{** the default language instead.}%
\else
\language=\csname l@#1\endcsname
\fi
#2}}
\providecommand{\BIBdecl}{\relax}
\BIBdecl
\renewcommand{\BIBentryALTinterwordstretchfactor}{4}

\bibitem{Graichen2007}
K.~Graichen, M.~Treuer, and M.~Zeitz, ``Swing-up of the double pendulum on a cart by feedforward and feedback control with experimental validation,'' \emph{Automatica}, vol.~43, no.~1, pp. 63--71, Jan. 2007.

\bibitem{Yamakita1995}
M.~Yamakita, M.~Iwashiro, Y.~Sugahara, and K.~Furuta, ``Robust swing up control of double pendulum,'' \emph{Proc. of 1995 Am. Control Conf.}, vol.~1, pp. 290--295, 1995.

\bibitem{Flasskamp2014}
K.~Flaßkamp, J.~Timmermann, S.~Ober-Blöbaum, and A.~Trächtler, ``Control strategies on stable manifolds for energy-efficient swing-ups of double pendula,'' \emph{Int. J. Control}, vol.~87, no.~9, 2014.

\bibitem{Fantoni2000}
I.~Fantoni, R.~Lozano, and M.~Spong, ``Energy based control of the pendubot,'' \emph{IEEE Trans. on Autom. Control}, vol.~45, no.~4, pp. 725--729, 2000.

\bibitem{Xin2012}
X.~Xin and T.~Yamasaki, ``Energy-based swing-up control for a remotely driven acrobot: Theoretical and experimental results,'' \emph{IEEE Trans. on Control Syst. Techn.}, vol.~20, no.~4, pp. 1048--1056, 2012.

\bibitem{Spong1995}
M.~Spong, ``The swing up control problem for the acrobot,'' \emph{IEEE Control Syst. Magazine}, vol.~15, no.~1, pp. 49--55, 1995.

\bibitem{Flasskamp2017}
K.~Flaßkamp, A.~R. Ansari, and T.~D. Murphey, ``Hybrid control for tracking of invariant manifolds,'' \emph{Nonlinear Analysis: Hybrid Syst.}, vol.~25, pp. 298--311, 2017.

\bibitem{Aastroem2000}
K.~Åström and K.~Furuta, ``Swinging up a pendulum by energy control,'' \emph{Automatica}, vol.~36, no.~2, pp. 287--295, 2000.

\bibitem{Shinbrot1992}
T.~Shinbrot, C.~Grebogi, J.~Wisdom, and J.~A. Yorke, ``Chaos in a double pendulum,'' \emph{Am. J. of Phys.}, vol.~60, no.~6, pp. 491--499, 1992.

\bibitem{AlbuSchaeffer2023}
A.~Albu-Sch{\"a}ffer and A.~Sachtler, ``What can algebraic topology and differential geometry teach us about intrinsic dynamics and global behavior of robots?'' in \emph{Robotics Research}, A.~Billard, T.~Asfour, and O.~Khatib, Eds.\hskip 1em plus 0.5em minus 0.4em\relax Springer Nature Switzerland, 2023, pp. 468--484.

\bibitem{Kerschen2009}
G.~Kerschen, M.~Peeters, J.~Golinval, and A.~Vakakis, ``Nonlinear normal modes, part i: A useful framework for the structural dynamicist,'' \emph{Mech. Syst. Signal Process.}, vol.~23, no.~1, pp. 170--194, 2009.

\bibitem{AlbuSchaeffer2020}
A.~Albu-Schäffer and C.~{Della Santina}, ``A review on nonlinear modes in conservative mechanical systems,'' \emph{Annu. Rev. Control}, vol.~50, pp. 49 -- 71, 2020.

\bibitem{Santina2021}
C.~{Della Santina} and A.~Albu-Schäffer, ``Exciting {Efficient} {Oscillations} in {Nonlinear} {Mechanical} {Systems} {Through} {Eigenmanifold} {Stabilization},'' \emph{IEEE Control Syst. Lett.}, vol.~5, no.~6, pp. 1916--1921, 2021.

\bibitem{Bjelonic2022}
F.~Bjelonic, A.~Sachtler, A.~Albu-Schäffer, and C.~Della~Santina, ``Experimental closed-loop excitation of nonlinear normal modes on an elastic industrial robot,'' \emph{IEEE Robot. Autom. Lett.}, vol.~7, no.~2, pp. 1689--1696, Apr. 2022.

\bibitem{Calzolari2023}
D.~Calzolari, C.~D. Santina, A.~M. Giordano, A.~Schmidt, and A.~Albu-Schäffer, ``Embodying quasi-passive modal trotting and pronking in a sagittal elastic quadruped,'' \emph{IEEE Robot. Autom. Lett.}, vol.~8, no.~4, pp. 2285--2292, 2023.

\bibitem{Sesselmann2021}
A.~Sesselmann, F.~Loeffl, C.~D. Santina, M.~A. Roa, and A.~Albu-Schäffer, ``Embedding a nonlinear strict oscillatory mode into a segmented leg,'' in \emph{Proc. IEEE/RSJ Int. Conf. Intell. Robots and Syst.}, 2021, pp. 1370--1377.

\bibitem{Sachtler2022}
A.~Sachtler and A.~Albu-Schäffer, ``Strict modes everywhere - bringing order into dynamics of mechanical systems by a potential compatible with the geodesic flow,'' \emph{IEEE Robot. Autom. Lett.}, vol.~7, no.~2, pp. 2337--2344, 2022.

\bibitem{Dickinson1976}
R.~P. Dickinson and R.~J. Gelinas, ``Sensitivity analysis of ordinary differential equation systems—a direct method,'' \emph{J Comp. Phys.}, vol.~21, no.~2, pp. 123--143, 1976.

\bibitem{Wotte2022}
Y.~P. Wotte, A.~Sachtler, A.~Albu-Schäffer, and C.~{Della Santina}, ``Sufficient conditions for an eigenmanifold to be of the extended {R}osenberg type,'' \emph{Research Square Preprint}, 2022.

\bibitem{Rosenberg1966}
R.~Rosenberg, ``On nonlinear vibrations of systems with many degrees of freedom,'' in \emph{Adv. in Appl. Mech.}, 1966, vol.~9, pp. 155 -- 242.

\bibitem{Barber1996}
C.~B. Barber, D.~P. Dobkin, and H.~Huhdanpaa, ``The quickhull algorithm for convex hulls,'' \emph{ACM Trans. Math. Softw.}, vol.~22, no.~4, p. 469–483, 1996.

\bibitem{Floater2015}
M.~S. Floater, ``Generalized barycentric coordinates and applications,'' \emph{Acta Numerica}, vol.~24, p. 161–214, 2015.

\bibitem{Sussman2001}
G.~J. Sussman and J.~Wisdom, \emph{Structure and Interpretation of Classical Mechanics}, 2nd~ed.\hskip 1em plus 0.5em minus 0.4em\relax Cambridge, USA: The MIT Press, 2015.

\bibitem{Strogatz2000}
S.~H. Strogatz, \emph{Nonlinear Dynamics and Chaos: With Applications to Physics, Biology, Chemistry and Engineering}, 2nd~ed.\hskip 1em plus 0.5em minus 0.4em\relax CRC Press, Taylor \& Francis Group, 2014.

\bibitem{Todorov2012}
E.~Todorov, T.~Erez, and Y.~Tassa, ``{MuJoCo: A} physics engine for model-based control,'' in \emph{Proc. IEEE/RSJ Int. Conf. Intell. Robots Syst.}, 2012, pp. 5026--5033.

\end{thebibliography}

\end{document}